\documentclass[]{article}
\usepackage{times,graphicx,a4wide}

\begin{document}
	
\title{Can Collimated Extraterrestrial Signals be Intercepted?}
\author{Duncan H. Forgan$^1$}
\maketitle

\begin{center}

$^1$Scottish Universities Physics Alliance (SUPA),\\School of Physics and Astronomy, University of St Andrews \\

\end{center}

\noindent \textbf{Word Count: 2877} \\

\noindent \textbf{Direct Correspondence to:} \\
D.H. Forgan \\
\textbf{Email:} dhf3@st-andrews.ac.uk \\

\newpage

\begin{abstract}

The Optical Search for Extraterrestrial Intelligence (OSETI) attempts to detect collimated, narrowband pulses of electromagnetic radiation.  These pulses may either consist of signals intentionally directed at the Earth, or signals between two star systems with a vector that unintentionally intersects the Solar System, allowing Earth to intercept the communication.  But should we expect to be able to intercept these unintentional signals? And what constraints can we place upon the frequency of intelligent civilisations if we do?

We carry out Monte Carlo Realisation simulations of interstellar communications between civilisations in the Galactic Habitable Zone (GHZ) using collimated beams.  We measure the frequency with which beams between two stars are intercepted by a third.    The interception rate increases linearly with the fraction of communicating civilisations, and as the cube of the beam opening angle, which is somewhat stronger than theoretical expectations, which we argue is due to the geometry of the GHZ.  We find that for an annular GHZ containing 10,000 civilisations, intersections are unlikely unless the beams are relatively uncollimated.

These results indicate that optical SETI is more likely to find signals deliberately directed at the Earth than accidentally intercepting collimated communications.  Equally, civilisations wishing to establish a network of communicating species may use weakly collimated beams to build up the network through interception, if they are willing to pay a cost penalty that is lower than that meted by fully isotropic beacons.  Future SETI searches should consider the possibility that communicating civilisations will attempt to strike a balance between optimising costs and encouraging contact between civilisations, and look for weakly collimated pulses as well as narrow-beam pulses directed deliberately at the Earth.

\end{abstract}


\newpage

\section{Introduction}

\noindent The Search for Extraterrestrial Intelligence (SETI) began in the radio bands of the electromagnetic (EM) spectrum.  The pervasive 21cm emission line of neutral atomic hydrogen happens to fall in the so-called ``Water Hole'' in the radio noise spectrum.  Given the innate scientific interest of the 21cm line, SETI scientists concluded that technological civilisations intending to communicate over interstellar distances might use radio transmitters tuned to frequencies on or around 21cm to attract the attention of other civilisations surveying the most common element in the Universe \cite{Morrison1977}.  This line of reasoning spawned a large number of SETI searches, from the initial ``blind'' surveys conducted by Drake \cite{Drake1961} to modern day searches, which target systems known to host extrasolar planets using single-dish \cite{Siemion2013} and interferometric techniques \cite{Rampadarath2012}.  

Radio searches focus on two broad classes of signal: intentional and unintentional.  Intentional signals can either be uncollimated ``beacons" designed to broadcast low information signals isotropically to any nearby listeners, or collimated ``beams'' designed to broadcast at specific targets.  Precisely which type of signal SETI scientists might expect depends on the motives of the broadcasting civilisation.  If cost is a controlling factor, then signals are likely to be collimated pulses \cite{Benford2010a,Benford2010, Messerschmitt2013}.  

Unintentional signals produced by humanity result from ``signal leakage'' from radio transmitters designed for terrestrial broadcasts.  As the leakage is ill-fit to travel large distances, it is difficult to detect except at close quarters, Galactically speaking, and relies on the civilisation not optimising their local communications to reduce this leakage \cite{SKA}.

Along a separate line of reasoning, Schwartz and Townes \cite{Schwartz1961} considered an alternative philosophy to transmitting and receiving messages at interstellar distances.  Contemporary developments in laser technology motivated the discussion of sending highly collimated beams of optical EM radiation.  Again, collimated pulses of narrowband optical radiation are cost-optimal for a fixed wavelength, and they possess several advantages over equivalent broadcasts at radio wavelengths.

The high gain of optical telescopes allows construction of beams that are much more collimated than those at longer wavelengths.  Dispersion of the signal due to the interstellar medium (ISM) is negligible at optical wavelengths \cite{Cordes2002}, and if measurements are made using photon counters operating at nanosecond timescales, there are very few sources of natural or man-made interference that can emit at sufficiently high intensity to confuse with an extraterrestrial signal \cite{Howard2001}.  Thanks to Moore's Law, instrumentation for optical SETI searches has become quite cheap relative to radio SETI, and is well-suited to piggyback operation on 1-metre class telescopes and above (see \cite{Siemion2011} for examples of both radio and optical SETI instrumentation).  Howard et al \cite{Howard2004} observed over 6000 stars using two geographically separated photon counters to eliminate background events and help isolate a putative pulse signal.  Of the roughly 300 candidate events, all were discounted as background events for various reasons.

While collimated pulses are the most cost-effective, it is possible that civilisations may use continuous laser beams to communicate, in effect adding an anomalous emission line to stellar spectra.  Reines and Marcy \cite{Reines2002} demonstrate that civilisations communicating in this mode can be detectable even if the laser beam power is as low as 50 kilowatts, provided that a sufficiently similar twin star can be found to robustly subtract an unmodified spectrum.  After studying the spectra of 577 nearby main sequence stars (with around 20 spectra per star collected over 4 years), they detect a single candidate line, which was discovered to be caused by a flaw in the CCD.  Continuous beams also offer the advantage of being easier to detect in archive data \cite{Borra2012}.

As with radio SETI, optical SETI searches can target two types of signal: intentional and unintentional.  In this case, an unintentional signal refers to a highly collimated signal sent between two civilisations, with a vector that happens to intersect a third civilisation's location: in this case, the Solar System. Given our ability to collimate optical signals (lasers) to very small solid angles, this leads us to the key question this paper intends to address: can we expect SETI searches to be successful at intercepting collimated signals not intended for humanity?

We attempt to answer this using Monte Carlo Realisation techniques.  We generate a Galactic Habitable Zone (GHZ) \cite{GHZ,Gowanlock2011} populated with stars hosting intelligent civilisations, and allow a fraction to emit collimated signals at a specific target star.  The number of intercepted signals is recorded as a function of time, beam opening angle and the fraction of communicating civilisations.  In section \ref{sec:methods} we describe the setup of the Monte Carlo Realisation model; in section \ref{sec:results} we describe and discuss the data arising from this model, and in section \ref{sec:conclusions} we summarise the work.


\section{Method} \label{sec:methods}

\noindent We utilise Monte Carlo Realisation techniques, which are commonly used in numerical SETI studies \cite{Vukotic_and_Cirkovic_07,Vukotic_and_Cirkovic_08,mcseti2,Forgan2011b, Nicholson2013}.  By conducting multiple simulations of a desired scenario, uncertainties in the model can be characterised easily.


\noindent In our analysis, each simulation generates a Galactic Habitable Zone (GHZ) of $N_*$ stars with intelligent civilisations, where we consider the GHZ as an annulus with inner and outer radii of 6 and 10 kiloparsecs (kpc) respectively.  Each star in the GHZ is assigned a set of Keplerian orbital elements, which are fixed throughout the simulation.  The semimajor axes of the stars $a_i$ are exponentially distributed to simulate the Milky Way's surface density profile:

\begin{equation} 
P(a_i) \propto e^{-\frac{a_i}{r_S}}
\end{equation}

with the scale radius $r_S=3.5$ kpc. The eccentricity distribution is uniform, under the constraint that a star's closest approach to the Galactic Centre must not be smaller than the inner radius of the GHZ.  We also restrict the inclination of the orbits so that they do not exceed 0.5 radians.  This is a relatively lax restriction, as this permits civilisations to approach altitudes of a few kpc above the midplane, but it does not affect our conclusions.  The longitude of the ascending node (and the argument of periapsis) are uniformly sampled in the range $[0,2\pi]$ radians.

A fraction $f_c$ of $N_*$ are selected to be the source of a communications beam of opening angle $\theta$ at a single target star, with a beam length $L$ equal to the separation of source and target.  Note that this implies the total number of \emph{beams} is equal to $f_cN_*$ - the number of communicating civilisations ranges between $f_cN_*$ and $2f_cN_*$ (where the maximum cannot exceed $N_*$), as we permit stars to be the target of multiple beams.  

\begin{figure}
\begin{center}
\includegraphics[scale = 0.4]{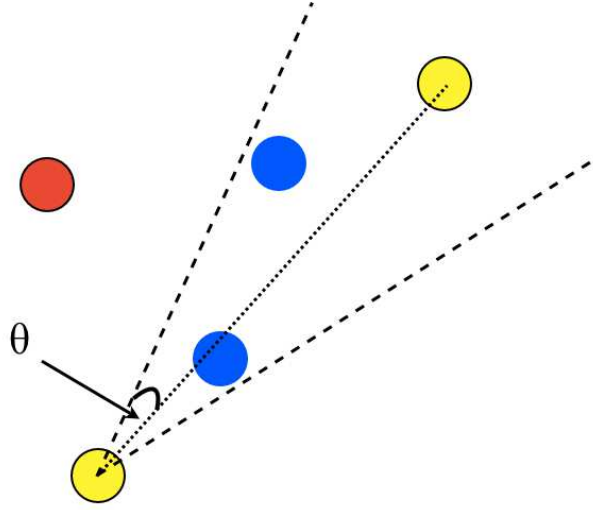}
\caption{Schematic of interception of collimated signals.  The source star emits a beam of fixed solid angle directed at the centre of the target star. Any star within the signal cone is intersected, regardless of distance. \label{fig:intersect_diagram}} 
\end{center}
\end{figure}

The stars are then evolved on their fixed orbits for 4.5 Gyr, with a simulation timestep of $10^{-3}$ Gyr. The beam's fixed solid angle (determined by the beam opening angle $\theta$) ensures that the cone of intersection increases from the source to the target (see Figure \ref{fig:intersect_diagram}).  Any stars that fall within this cone (and are neither the target or the source) are regarded as intercepting a communication.  At any time $t$, we can therefore calculate the total number of interceptions by all stars, $n_i(t)$ (including stars that are sources or targets of other beams).


The Monte Carlo Realisation procedure can be summarised as follows.  For each run:

\begin{enumerate}
\item Generate $N_*$ stars, with fixed Keplerian orbital elements
\item Select $f_cN_*$ source stars, and give each source a single target.
\item Simulate the motion of stars over multiple timesteps.  At every timestep, record $n_i(t)$.
\end{enumerate}

\noindent This process is repeated 30 times with different random number seeds, to produce 30 different realisations of $n_i(t)$.  We calculate the mean intersection count over all realisations, $\bar{n}_i(t)$, and then average this quantity over time to obtain $\bar{n}$, which is the expected number of intersections inside the GHZ at any time over the last 4.5 Gyr.   We also obtain the standard error of the mean for $\bar{n}$, which we label $\sigma_n$.

\section{Results \& Discussion} \label{sec:results}

\subsection{Expected Trends \label{sec:expected_trends}}

\noindent Before running this numerical experiment, we can make some simple calculations to assess what trends we might expect to see in the data.  The expected number of intersections can be expressed as:

\begin{equation}
\bar{n} = N_{beam} <{V}_{beam}> \rho_{civ},
\end{equation}

\noindent where $N_{beam}$ is the number of beams, $<V_{beam}>$ is the mean volume enclosed by each beam, and $\rho_{civ}$ is the number density of civilisations (which we will take to be uniform for now).  If we substitute $<V_{beam}>$ for the  volume of a cone with mean length $<L>$, and express the other properties in terms of $f_c$ and $N_*$:

\begin{equation}
\bar{n}  = f_c N_* \frac{\pi <L^3> \theta^2}{3} \frac{N_*}{V_{GHZ}},
\end{equation}

\noindent where $V_{GHZ}$ is the volume of the GHZ.  In other words, we should expect

\begin{equation}
\bar{n} \propto f_c N^2_* \theta^2,
\end{equation}

\noindent which we now test in the following sections.

\subsection{Dependence on Beam Solid Angle}

\begin{figure}
\begin{center}$
\begin{array}{cc}
\includegraphics[scale = 0.4]{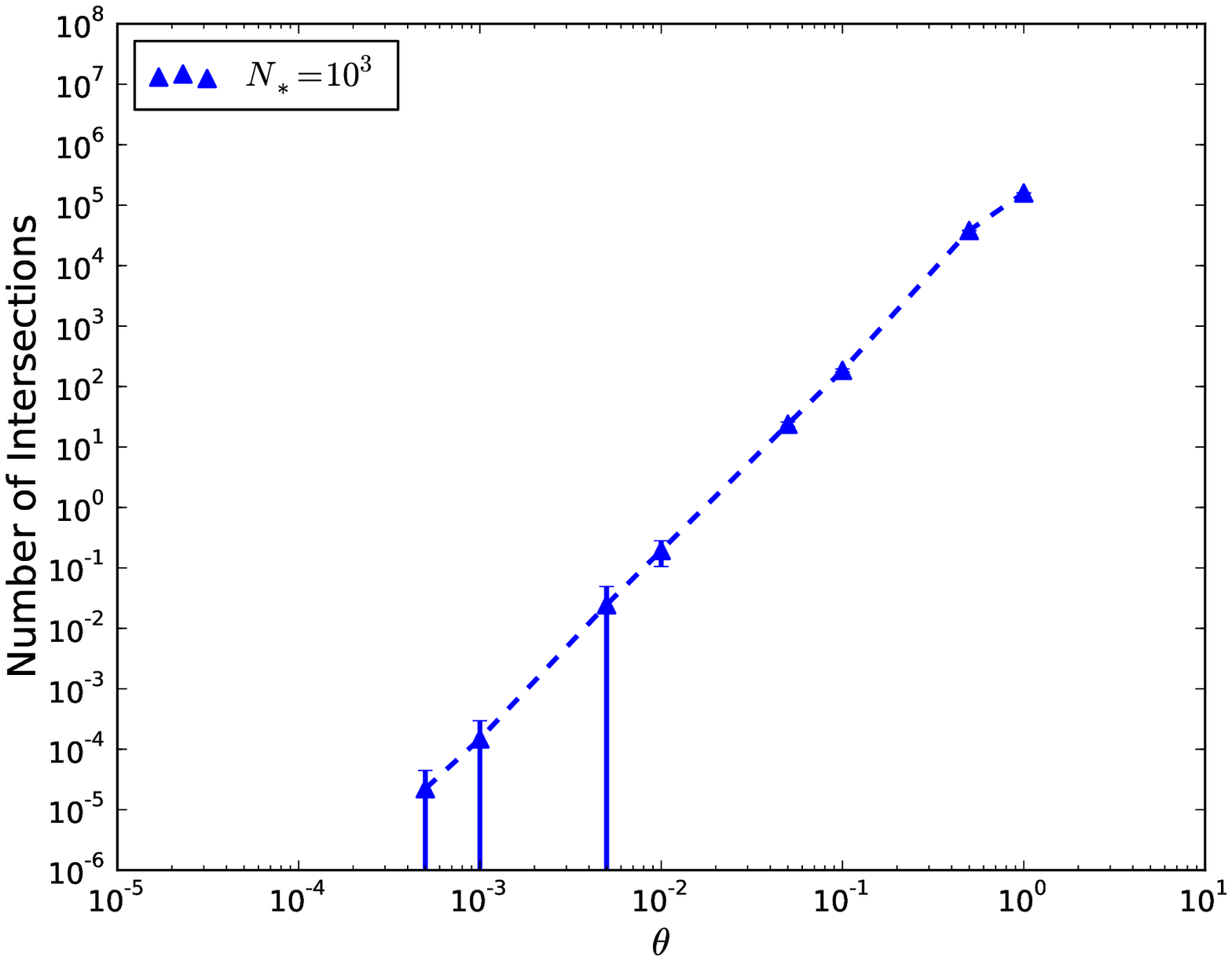} &
\includegraphics[scale = 0.4]{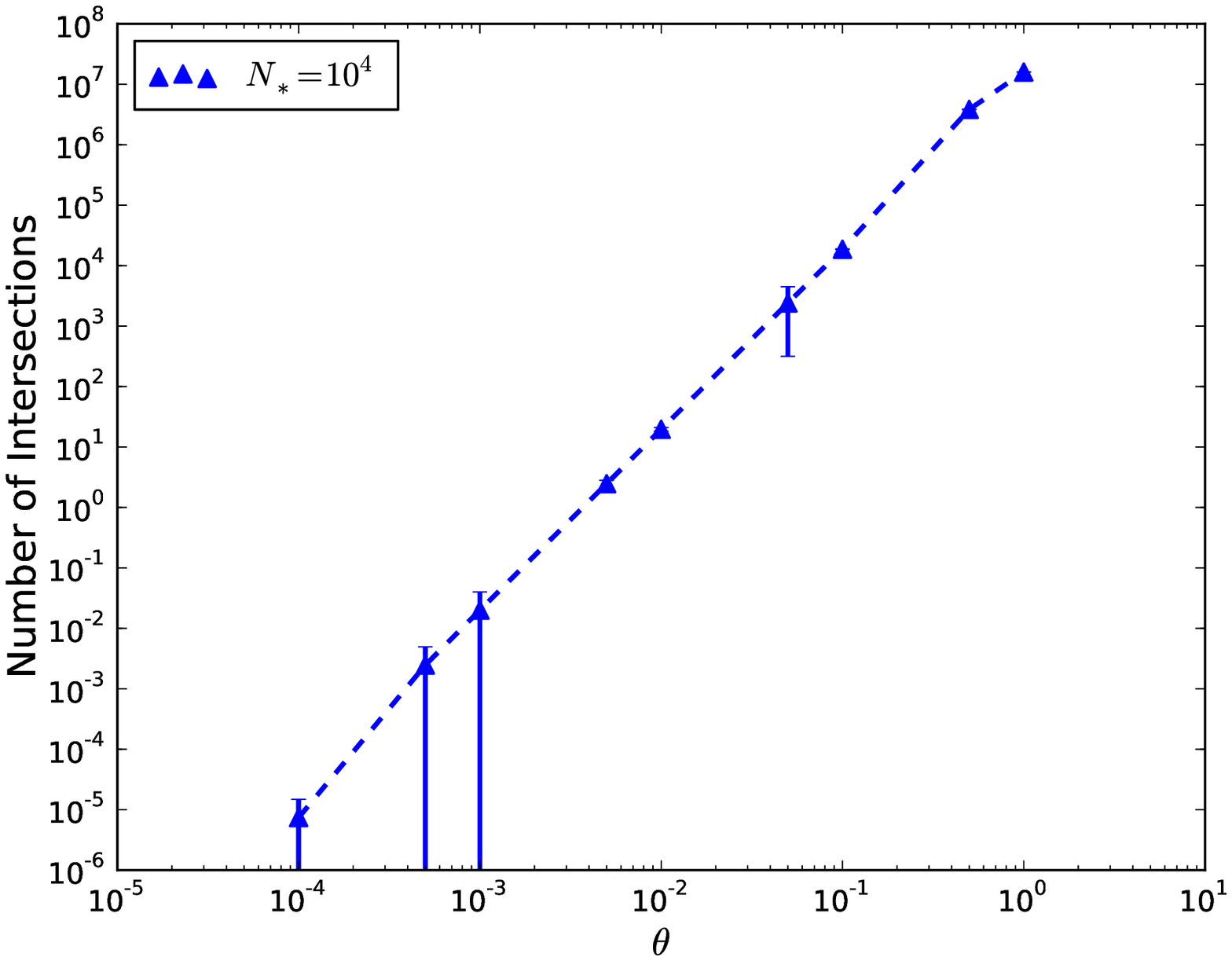} \\
\end{array}$
\caption{How the time averaged number of intersections varies with beam angle (for a fixed communication fraction $f_c=0.5$, for $N_*=10^3$ (left) and $N_*=10^4$ (right).  We also plot error bars indicating $\sigma_n$ for each point, which becomes larger than $\bar{n}$ for $\bar{n}<1$. \label{fig:nintersect_vs_omega}} 
\end{center}
\end{figure}

\noindent Firstly, we consider a GHZ with $N_*$ stars hosting civilisations, and communication fraction $f_c=0.5$, and vary the beam opening angle $\theta$.  In Figure \ref{fig:nintersect_vs_omega} we display $\bar{n}$ as a function of $\theta$ (with error bars corresponding to $\sigma_n$) for $N_*=10^3$ and $N_* = 10^4$.  In both cases, we find a simple powerlaw relationship between the intersection number and $\theta$: 

\begin{equation}
\bar{n} \propto \theta^3
\end{equation}

\noindent The intersection number attains its maximum for $\theta = \pi/2$, where each beam covers the entire sky and is intercepted by all stars in the GHZ (with the exception of the source and target stars, by our definition of interception):

\begin{equation}
n_{i,max} = (f_c N_*) (N_* - 2)
\end{equation}

\noindent When $\bar{n}<1$ at small values of $\theta$, the standard error $\sigma_n$ becomes relatively large, giving the large errors we see in Figure \ref{fig:nintersect_vs_omega}. 

We can see that in both cases for beam angles below a critical value, $\theta_{crit}$, the typical number of interceptions any civilisation may expect to make at any given timestep, $\bar{n}/N_*$, averages to less than 1 .  The value of $\theta_{crit}$ depends on the volume density of interceptors (as well as $f_c$).  For $N_*=10^3$, $\theta_{crit} \approx 0.2$ rad, and for $N_*=10^4$, $\theta_{crit} \approx 0.08$ rad.   This is about 1-2 orders of magnitude larger than a commercial laser pointer, and significantly larger than the proposed opening angles of order 10 mas for the ``Earth 2000'' communication scenario proposed by Howard et al \cite{Howard2004}.

\subsection{Dependence on Communication Fraction}

\begin{figure*}
\begin{center}$
\begin{array}{cc}
\includegraphics[scale = 0.4]{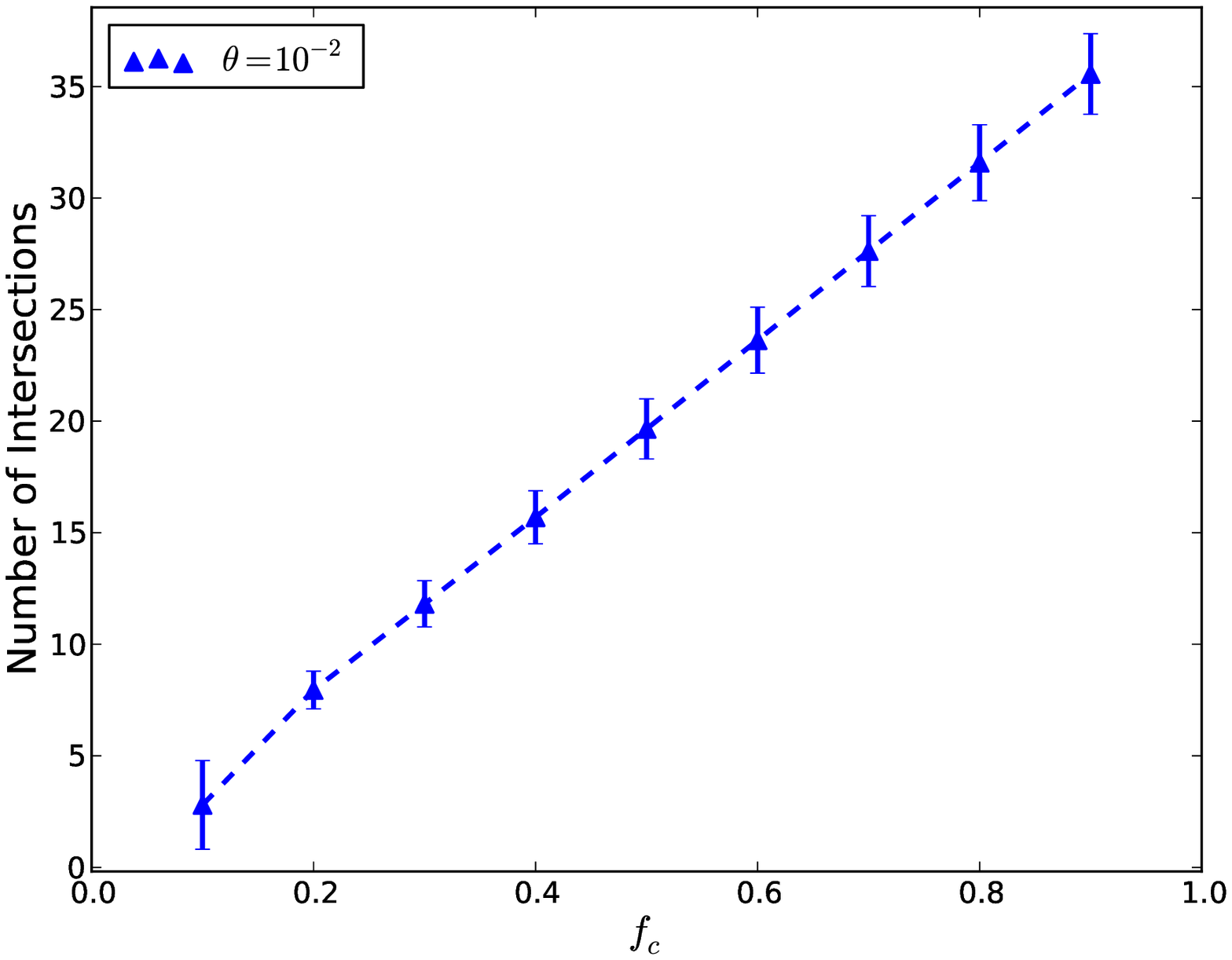} &
\includegraphics[scale=0.4]{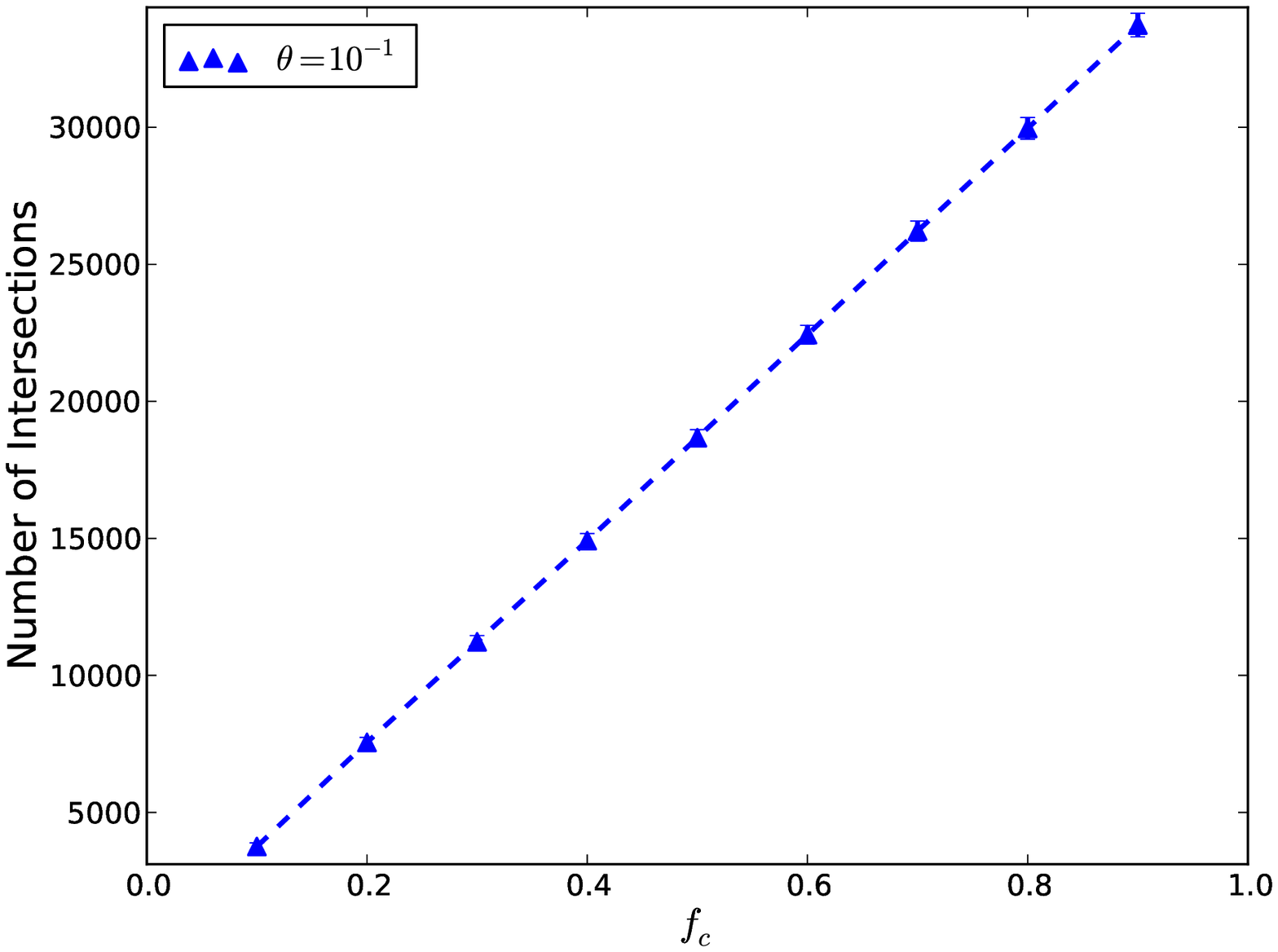} \\
\end{array}$
\caption{The variation of the time averaged number of intersections with communication fraction $f_c$, for two fixed beam angles: $\theta=10^{-2}$ (left) and $\theta=10^{-1}$ (right). \label{fig:nintersect_vs_fc}} 
\end{center}
\end{figure*}

\noindent We now consider how $\bar{n}$ depends on $f_c$, for two relatively large values of $\theta = 10^{-2},10^{-1}$ (Figure \ref{fig:nintersect_vs_fc}).  Generally, the interception number increases linearly with $f_c$, independent of $\theta$, and is well fitted by 

\begin{equation}
\bar{n} = C f_c
\end{equation}

\noindent Where $C$ also depends on $\theta$: for $\theta=10^{-2}$, $C \approx 39$, and for $\theta=10^{-1}$, $C\approx 3.7\times10^{4}$.  

\subsection{General Trends}

We therefore propose the following trend:

\begin{equation}
\bar{n} \propto f_c N_*^2 \theta^3,
\end{equation} 

\begin{figure*}
\begin{center}$
\begin{array}{cc}
\includegraphics[scale = 0.4]{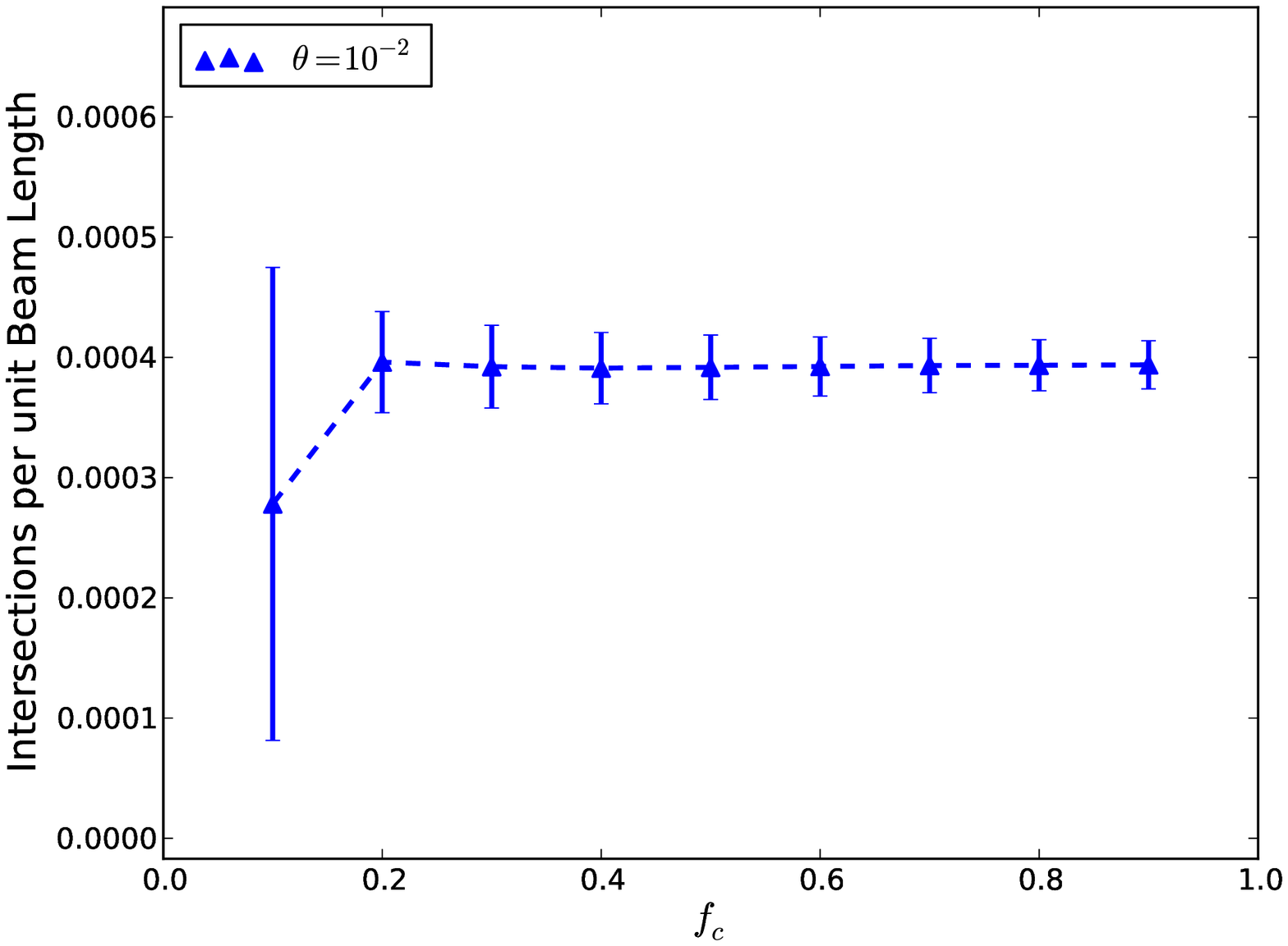} &
\includegraphics[scale=0.4]{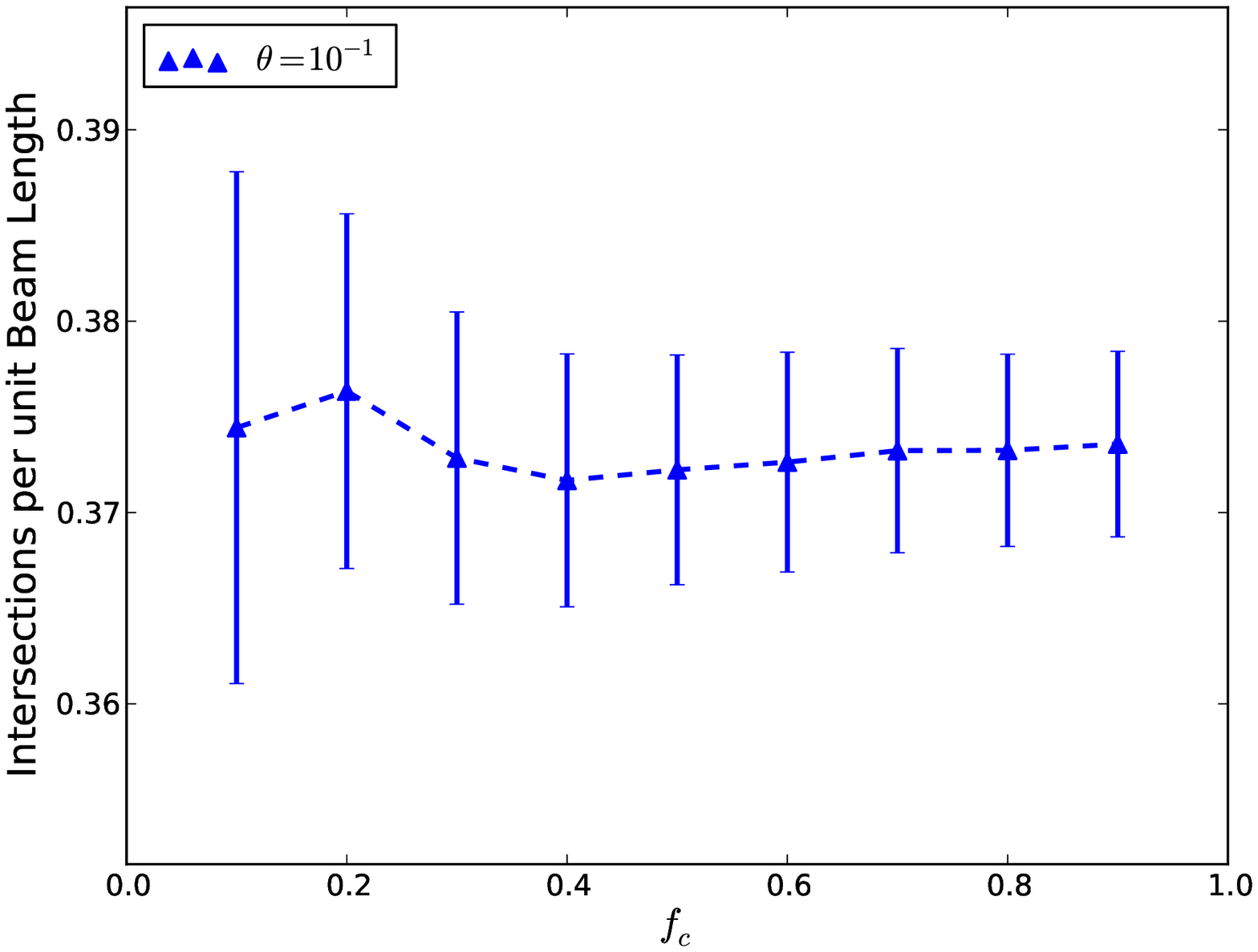} \\
\end{array}$
\caption{The time averaged number of intersections per unit beam length as a function of communication fraction $f_c$, for two fixed beam angles: $\theta=10^{-2}$ (left) and $\theta=10^{-1}$ (right). \label{fig:nperbeam_vs_fc}} 
\end{center}
\end{figure*}

\noindent Where the constant of proportionality depends on the geometric properties of the GHZ.  We can confirm this by investigating the average number of intersections per unit beam length, $\bar{n}/L$ (Figure \ref{fig:nperbeam_vs_fc}).  This quantity can be calculated by dividing $\bar{n}$ by the typical total length of all beams at any timestep.  The total beam length will depend linearly on the number of beams ($f_cN_*$), and as such we should therefore expect $\bar{n}/L$ to be independent of $f_c$, which is borne out by the data - for both values of $\theta$, the behaviour of $\bar{n}/L$ against $f_c$ is consistent with a straight line of gradient zero.

\noindent Is this form for $\bar{n}$ sensible?   It is close to the trend we calculated in section \ref{sec:expected_trends}, with the exception of the $\theta$ dependence.  However, our original calculation assumed a homogeneous, isotropic GHZ, and our simulations use an annular GHZ with a number density that decreases exponentially in radius, which we argue explains the slightly steeper dependence on beam angle.

\subsection{Limitations of the Analysis}

\noindent We have deliberately simplified this analysis to uncover broad trends in interception of collimated signals, and we note there are several extra considerations that may affect the interception rate.  Firstly, we have assumed that any two civilisations in the GHZ may establish communication via collimated beams.  This assumption is composed of two smaller assumptions:

\begin{enumerate}
\item Civilisations can communicate across any distance,
\item Civilisations exist throughout the 4.5 Gyr runtime of the simulation.
\end{enumerate}

\noindent The first requires that civilisations possess transmitter technology with sufficient emitting power to produce collimated signals that travel a distance $d$ that must be equal to twice the outer radius of the GHZ for all civilisations to be communicable (given  $\theta$).  For a civilisation to receive the signal, they must be able to detect signals of a minimum flux $F_{min}$.  If we take terrestrial laser technology as a template, then the typical beam power is of order a petawatt (where we assume a 3 MJ pulse of duration 3 nanoseconds) and transmission distances of several kpc are feasible \cite{Howard2004}, which is somewhat less than our GHZ's value for $d=20$ kpc, despite being well-collimated.  

The second of the above assumptions (that civilisations exist for the full 4.5 Gyr) is likely to be incorrect, as we expect civilisations to emerge and disappear at different times, with these timescales correlated with star formation and planet formation epochs \cite{Vukotic_and_Cirkovic_07,mcseti1, Gowanlock2011}.  Adding a temporal constraint to communications can significantly reduce the ability of civilisation pairs to communicate directly, much less intercept communications \cite{mcseti2}.   

We have also assumed that each civilisation only communicates to one other, i.e. that communication is exclusively pairwise.  It seems rational that communication networks will develop as civilisations discover each other, either through detecting uncollimated or collimated transmissions, or unintentional signals such as leakage.  

Indeed, we might expect that once a civilisation intercepts a collimated beam between civilisations, the interceptor may attempt to establish communication with either the source, the target, or both.  Interceptions do appear to be relatively rare compared to the number of beams, but the interception rate will increase as more beams come online.  If the rate of increase of interceptions depends on the rate of increase of beams:

\begin{equation} 
\dot{n} = \dot{f}_c,
\end{equation}

\noindent This differential equation produces a solution for $\bar{n}$ which is exponentially growing, where the growth factor is sensitive to $\theta$.  A group of civilisations attempting to foster contact and develop a network of contacted species may tailor their beams to encourage interception, and subsequent addition to the network.  While this significantly increases the cost of transmission compared to collimated signals \cite{Benford2010}, it is far lower than the cost of using isotropic beacons, with the added benefit of maintaining communication with known civilisations.

Finally, we have assumed that the civilisations are distributed in a GHZ between 6 and 10 kpc.  This is modelled on calculations that show the GHZ to be more extended \cite{Gowanlock2011} compared to the classic calculations by Lineweaver et al \cite{GHZ} which established the GHZ concept.  We do not expect the interception rate to increase much by altering the precise configuration of the GHZ, although it may introduce a slightly different dependence on $\theta$.


\section{Conclusions}\label{sec:conclusions}

\noindent We have conducted Monte Carlo Realisation simulations of collimated signal communication between civilisation pairs in the Galactic Habitable Zone (GHZ), in an attempt to understand the likelihood of those signals being intercepted by other civilisations.  We find that the number of intersections at any time scales as the number of civilisations squared, and the cube of the beam opening angle, which is slightly steeper than simple theoretical calculations predict (and appears to be due to the geometry of the GHZ).  

The number of intersections is relatively low compared to the number of civilisations, unless the beam opening angle is relatively large (of order 0.1 radians).  This suggests that SETI searches are unlikely to find signals that are highly collimated and not aimed at the Earth (unless the number of communicating civilisations is very large).

This still leaves several possibilities for the detection of extraterrestrial signals by humanity:

\begin{enumerate}
\item Highly collimated signals sent deliberately at the Earth, 
\item Uncollimated or isotropic beacons designed to contact hitherto uncontacted civilisations, and
\item Weakly collimated signals sent between civilisation pairs, designed to encourage interception by unconnected civilisations like ours, while maintaining a transmission cost that is low compared to isotropic beacons.
\end{enumerate}

\noindent All require civilisations to desire contact with previously uncontacted civilisations, either directly or indirectly.  In short, optical SETI remains a valid component of the SETI programme, but it is clear that searches of this type must be realistic as to what types of signal they are likely to detect.  Current SETI programmes are quite sensitive to option 1 above, but the weaker signals of option 3 may be more common depending on the motives of civilisations, especially if they wish to use interception to quickly build a highly connective network of civilisations without explicitly searching for them, or using large amounts of energy to build beacons.  We recommend that future instrumentation and search designs consider this possibility.


\section*{Acknowledgments}

\noindent The author gratefully acknowledges support from STFC grant ST/J001422/1.

\bibliographystyle{abbrv} 
\bibliography{lasercomm_intercept}

\end{document}